\documentclass[twocolumn,showpacs,preprintnumbers]{revtex4}
\usepackage{graphicx}
\usepackage{dcolumn}
\usepackage{bm}

\newcommand{\be}{\begin{equation}}
\newcommand{\beq}{\begin{eqnarray}}
\newcommand{\en}{\end{equation}}
\newcommand{\enq}{\end{eqnarray}}

\begin{document}
\topmargin -1cm
\title{Radiation in (2+1)-dimensions}
\author{Mauricio Cataldo}
\altaffiliation{mcataldo@ubiobio.cl}
\affiliation{Departamento de F\'\i sica, Facultad de Ciencias,
Universidad del B\'\i o-B\'\i o, Avenida Collao 1202, Casilla 5-C,
Concepci\'on, Chile.\\}
\author{Alberto A. Garc\'\i a}
\altaffiliation{aagarcia@fis.cinvestav.mx}
\affiliation{Departamento~de~F\'{\i}sica.
\\~Centro~de~Investigaci\'on~y~de~Estudios~Avanzados~del~IPN.\\
Apdo. Postal 14-740, 07000 M\'exico DF, MEXICO. \\}

\date{\today}

\begin{abstract}
In this paper we discuss the radiation equation of state $p=\rho/2$
in (2+1)-dimensions. In (3+1)-dimensions the equation of state
$p=\rho/3$ may be used to describe either actual electromagnetic
radiation (photons) as well as a gas of massless particles in a
thermodynamic equilibrium (for example neutrinos). In this work it
is shown that in the framework of (2+1)-dimensional Maxwell
electrodynamics the radiation law $p=\rho/2$ takes place only for
plane waves, i.e. for $E = B$. Instead of the linear Maxwell
electrodynamics, to derive the (2+1)-radiation law for more general
cases with $E \neq B$, one has to use a conformally invariant
electrodynamics, which is a 2+1-nonlinear electrodynamics with a
trace free energy-momentum tensor, and to perform a volumetric
spatial average of the corresponding Maxwell stress-energy tensor
with its electric and magnetic components at a given instant of time
$t$. 
\end{abstract}
\maketitle \preprint{APS/123-QED}

\section{Introduction}

It is well known that radiation or black body radiation (as a
superposition of plane waves of different frequencies) from the
point of view of a perfect fluid obeys the equation of state
$p=\rho/3$. Additionally, there are massless particles which in the
standard framework may be treated in terms of a fluid with energy
density $\rho$ and isotropic pressure $p$, which satisfies the same
equation of state. This law of radiation has been established in the
theory of gases, in particular, by means of the virial
theorem~\cite{Landau,Lightman}. The virial theorem to describe
radiation of electromagnetically interacting ultra-relativistic
particles has been used in Ref.~\cite{Landau} (ch. 5), where is
pointed out that one arrives at the $1/3$-radiation law since the
Maxwell energy-momentum tensor is characterized by the vanishing of
its trace. This property plays a crucial role in establishing the
correspondence ``gas-particle-field". In 3+1 dimensions within
nonlinear electrodynamics there is no way to establish the quoted
radiation law by averaging the corresponding energy-momentum tensor
with non-vanishing trace.

To obtain in (2+1)-spacetime the radiation law $p=\rho/2$, which can
be established from the corresponding formal gas thermodynamics,
from the view point of electrodynamics, one has to construct an
energy-momentum tensor by means of the electrodynamics we proposed
previously~\cite{Cataldo}. If one were averaging the (2+1)-(linear)
Maxwell energy-momentum tensor $T_{\mu\nu}$ one never should obtain
the relation $p=\rho/2$ for ultra-relativistic particles interacting
via this electrodynamics. This fact compelled us to search for an
equivalent to the (3+1)-radiation formulation.

The main goal of this paper is just to establish a parallelism
between radiation equation of states $p=\rho/3$ and $p=\rho/2$  in
(3+1) and (2+1)-gravities respectively.

From Cosmology, described by the Friedmann-Robertson-Walker (FRW)
model,
\begin{eqnarray}
ds^2=dt^2 -a(t)^2 \left(\frac{dr^2}{1-kr^2}+r^2(d \theta^2 +\sin^2
\theta d \varphi^2) \right),
\end{eqnarray}
where $k=-1,0,1$ and $a(t)$ is the scale factor, one can easily
obtain the $1/3$-state equation for radiation: In this model all
distances are proportional to the scale factor $a(t)$, thus the
volume $V$ scales as $a(t)^{3}$, and the wavelength $\lambda$ of
electromagnetic waves is proportional to $a(t)$, therefore the
corresponding radiation energy density of one photon in a volume $V$
is given by $\rho_{_{photon}}=\frac{h \nu}{V}=\frac{h c}{\lambda
V}$, this means that this energy density behaves like
$\rho_{_{photon}}= \rho_{_{0}} a(t)^{-4}$. On the other hand, using
the energy conservation equation of a perfect fluid,
\begin{eqnarray}
\dot{\rho}+ 3 \frac{\dot{a}}{a} (p+\rho)=0,
\end{eqnarray}
one derives the expression of the pressure in terms of the density,
which occurs to be just
\begin{eqnarray}\label{radiacion3+1}
p=\frac{1}{3} \, \rho,
\end{eqnarray}
and with the help of the Friedmann equation
\begin{eqnarray}\label{FE}
3 \left(\frac{\dot{a}}{a} \right)^2+\frac{3k}{a^2} = \kappa_4 \rho,
\end{eqnarray}
where $\kappa_4=8 \pi G$, we may find, for example, that for a flat
FRW model the scale factor is given by $a(t)=a_0 t^{1/2}$, where
$a_0$ is a constant.

In (2+1)-FRW cosmology
\begin{eqnarray}\label{2+1FRW}
ds^2=dt^2 -a(t)^2 \left(\frac{dr^2}{1-kr^2}+r^2 d \theta^2 \right),
\end{eqnarray}
one may accomplish a similar treatment for (2+1)-radiation. In this
case the radiation energy density $\rho_{_{photon}}=\frac{h
\nu}{S}=\frac{h c}{\lambda S}$ amounts to $\rho_{_{photon}} =
\rho_{_{0}} a(t)^{-3}$. Hence by using the energy conservation
equation of a perfect fluid,
\begin{eqnarray}
\dot{\rho}+ 2 \frac{\dot{a}}{a} (p+\rho)=0,
\end{eqnarray}
one obtains that the pressure fulfills
\begin{eqnarray}\label{stateeq3}
p=\frac{\rho}{2},
\end{eqnarray}
and with the help of the three-dimensional Friedmann equation
\begin{eqnarray}\label{2+1FE}
\left(\frac{\dot{a}}{a} \right)^2+\frac{k}{a^2} = \kappa_3 \rho,
\end{eqnarray}
we may find, for example, that for a flat FRW model the scale factor
is given by $a(t)=a_0 t^{2/3}$, where $a_0$ is a constant.

In Section II we briefly recall the averaging of the energy-momentum
tensor approach to radiation in (3+1)-dimensions. In Section III the
averaging  approach is applied to the (2+1)-maxwelian tensor and
also to the (2+1)-nonlinear electrodynamics tensor singled out by
the vanishing of its trace: in the Maxwell case one arrives at the
stiff matter state equation $p=\rho$, which is far from being the
$1/2$-radiation law, while for the energy-momentum tensor of a
conformally invariant electrodynamics, which is a 2+1-nonlinear
electrodynamics with vanishing trace, one obtains the radiation law
$p=\rho/2$. The Cornish and Frankel cosmological radiative solution
is commented and some concluding remarks are added.

\section{Radiation in (3+1)-dimensions}
Many years ago, in 1930, Tolman and Ehrenfest~\cite{Tolman} analyzed
the problem of the black body radiation in the framework of
thermodynamic equilibrium of a matter distribution described by a
perfect fluid energy-momentum tensor, using a static spherically
symmetric spacetime. These authors established the black body
radiation state equation~(\ref{radiacion3+1}) by treating this
phenomenon from Maxwell electrodynamics via an averaging procedure
of the components of the Maxwell energy-momentum tensor.

For the sake of reference, we repeat this procedure in details: the
electromagnetic field is specified at any point by the Maxwell
tensor $F_{\alpha \beta}$, and the Maxwell energy momentum-tensor is
given by
\begin{eqnarray}\label{TEMEM}
T_{\alpha \beta}=-F_{\alpha \gamma}F_{\beta}^{\,\,\,\,\, \gamma} +
\frac{1}{4} \, g_{\alpha \beta} \, F_{\gamma \delta} \, F^{\gamma
\delta},
\end{eqnarray}
where $F_{\gamma \delta}F^{\gamma \delta}=2 (B^2-E^2)$ is an
invariant of the electromagnetic field.

It is well known that electromagnetic fields are not compatible with
highly symmetric spacetimes, such as isotropic and homogeneous FRW
spacetimes, due to inherent anisotropic nature of Maxwell field
sources. This electromagnetic field can be included as a source of
such spacetimes through an averaging procedure, yielding then to an
effective perfect fluid source with an isotropic pressure. In order
to achieve the isotropy of the fields one has to require that the
electric and magnetic fields components do not possess preferred
directions thus the mean values fulfill the following
relations~\cite{Tolman}
\begin{eqnarray}
\overline{E_i}=\overline{B_i}=\overline{E_i \, B_i}=0,
\end{eqnarray}
and
\begin{eqnarray}
\overline{E_i \, E_j}=-\frac{1}{3} E^2 g_{ij}, \,\,\,\,\,
\overline{B_i \, B_j}=-\frac{1}{3} B^2 g_{ij},
\end{eqnarray}
where the bar over physical quantities stands for volumetric spatial
average of the corresponding quantities at a given instant of time
$t$.

\noindent Then, considering the metric $(+1,-1,-1,-1)$, one has that
$\overline{E^2_i}=E^2/3$, and $\overline{B^2_i}=B^2/3$. Using the
above average relations, one obtains by averaging the
energy-momentum tensor~(\ref{TEMEM}), the following perfect fluid
configuration
\begin{eqnarray}
T^{00}&=&\rho_{em}, \,\,\, T^{11}=T^{22}=T^{33}=p_{em},
\end{eqnarray}
where
\begin{eqnarray}\label{SEEMF}
p_{em}&=&\frac{1}{3}\rho_{em}=\frac{1}{6}(E^2+B^2),
\end{eqnarray}
hence the radiation state equation~(\ref{radiacion3+1}) takes place.

Note that the state equation~(\ref{SEEMF}) is not only valid for a
field of plane waves, for which $E=B$. In the deduction of
Eq.~(\ref{SEEMF}) we did not say anything about the properties of
the electromagnetic field, so the strengths of the electric and
magnetic fields may take any value. It may be in particular a
chaotic magnetic field (for which $E=0, B \neq 0$) or a random
magnetic field~\cite{Enqvist}.

Consequently, ``radiation'' may be used to describe either actual
electromagnetic radiation (for massless photons and neutrinos this
state equation is exactly valid), or massive particles moving at
relative velocities sufficiently close to the speed of light for
which the state equation~(\ref{radiacion3+1})  takes place
asymptotically. In this case due to that the velocities of the gas
particles approach that of light their rest energy becomes
negligible compared to their total energy. Thus by neglecting their
rest masses fluid behaves like electromagnetic radiation.

Although radiation is a perfect fluid and thus has an
energy-momentum tensor given by $T_{\mu
\nu}^{^{PF}}=(p+\rho)u_{\mu}u_{\nu}-p g_{\mu\nu}$ (with trace
$T^{^{PF}}=\rho-3p $), we also know that $T^{\mu\nu}$ can be
expressed in terms of the field strength~(\ref{TEMEM}). The trace of
this is given by $T^\mu{}_\mu =
{1\over{4\pi}}\left[F^{\mu\nu}F_{\mu\nu} -{1\over
4}(4)F^{\lambda\sigma}F_{\lambda\sigma}\right] = 0$. But this must
also equal $T^{^{PF}}=\rho-3p $, so the equation of state
is~(\ref{radiacion3+1}).

From here we deduce that the term ``radiation", or more exactly the
state equation~(\ref{radiacion3+1}), is used in a more general
sense. Effectively, a such interpretation of ``radiation" state
equation can be introduced into the study of cosmological models in
the early universe, where matter should be identified with a
primordial plasma. This is equivalent to put the squared electric
field $E^2=0$ in Eq.~(\ref{SEEMF}), neglecting bulk viscosity terms
in the electric conductivity of the primordial
plasma~\cite{Novello}.

Also one can study a universe filled with a non-interacting chaotic
or random magnetic field and radiation.

\section{Radiation in (2+1)-dimensions}
Now we shall treat the (2+1)-dimensional case. Usually one considers
the same Maxwell electromagnetic energy-momentum
tensor~(\ref{TEMEM}) to describe electromagnetic phenomena in
(2+1)-dimensions. In these dimensions the Maxwell tensor $F_{\alpha
\beta}$ has only three independent components, two for the vector
electric field $E_i$ and one for the magnetic field $B$, which now
is a pseudoscalar field, in contrast to four-dimensional Maxwell
field. Nevertheless, the use of the Maxwell electrodynamics in
different dimensions deserves some attention. From Eq.~(\ref{TEMEM})
we obtain that in a (N+1)-dimensional spacetime the energy-momentum
tensor trace is given by
\begin{eqnarray}\label{TRACE TEMEM}
T_{N+1}=T_{\alpha \beta} g^{\alpha \beta}=\left(-1 +\frac{N+1}{4}
\right) F_{\gamma \delta} F^{\gamma \delta}.
\end{eqnarray}
It becomes clear that in (3+1)-dimensions the electromagnetic tensor
has a vanishing trace; this property, being an invariant one,
singles out the Maxwell theory as the only trace free linear theory
in (3+1)-dimensions. Additionally, the propagation velocity of
electromagnetic waves coincides with the velocity of propagation of
the gravitational waves. Moreover, the eigenvalue problem, as it
should be, presents its own features depending on the
dimensionality. On the other hand, from Eq.~(\ref{TRACE TEMEM}) we
see that $T_{2+1}=-1/4 \, F_{\gamma \delta} F^{\gamma \delta}$,
hence in (2+1)-dimensions the Maxwell energy-momentum tensor
possesses a non-vanishing trace. As we shall see below, this implies
that, in the framework of Maxwell electromagnetism, the equation of
state $p=\rho/2$ takes place only for plane waves.

Notice that any comparison  of the velocity of the electromagnetic
waves with gravitational waves in (2+1)-dimensions is empty, since
there are no (vacuum) gravitational waves in these dimensions.

In what follows, we shall extend the well established averaging
procedure of the (3+1)-theory to the (2+1)-case for linear and
conformally invariant nonlinear electrodynamics.

\subsection{Maxwell electrodynamics; stiff matter and dust}
Assuming~(\ref{dosF}) in (2+1)-gravity that electrodynamics is
described by Maxwell theory with the energy-momentum
tensor~(\ref{TEMEM}), the averaging procedure yields
\begin{eqnarray}\label{8}
\overline{E_i}=\overline{E_i \, B}=0,
\end{eqnarray}
and
\begin{eqnarray}\label{9}
\overline{E_i \, E_j}=-\frac{1}{2} E^2 g_{ij}, \,\,\,\,\,
\overline{B^2}=B^2,
\end{eqnarray}
($\overline{E^2_i}=E^2/2$, for the metric $(+1,-1,-1)$). Hence the
averaging of the (2+1)-Maxwell electromagnetic energy-momentum
tensor yields
\begin{eqnarray}
T^{00}_{_{2+1}}&=&\rho^{em}_{_{2+1}},
T^{11}_{_{2+1}}=T^{22}_{_{2+1}}= p^{em}_{_{2+1}},
\end{eqnarray}
where
\begin{eqnarray}\label{11}
\rho^{em}_{_{2+1}}&=&\frac{1}{2}(E^2+B^2), \,\,\,\,\,
p^{em}_{_{2+1}}=\frac{1}{2} B^2.
\end{eqnarray}
Consequently, from Eqs~(\ref{11}), if $E=0$ one concludes that a
stiff matter state equation arises:
\begin{eqnarray}\label{2+1 B}
p^{em}_{_{2+1}}=\rho^{em}_{_{2+1}}= B^2/2,
\end{eqnarray}
which can be called plasma, in correspondence with the terminology
used in (3+1)-dimensions. In this case, by using the
three-dimensional Friedmann equation~(\ref{2+1FE}) with $k=0$, we
have that for a flat FRW model the energy density takes the form
$\rho(t)=\rho_0 a^{-4}$, i.e. $B \sim a^{-2}$ and the scale factor
is given by $a(t)=a_0 t^{1/2}$, where $a_0$ is a constant.

\noindent Next, if $B=0$ the matter distribution can be viewed as
dust:
\begin{eqnarray}\label{2+1 E}
\rho^{em}_{_{2+1}}=E^2/2, \,\,\,\,\,\,\,\,\,\, p^{em}_{_{2+1}}=0.
\end{eqnarray}
In this case, by using Eq.~(\ref{2+1FE}) with $k=0$, we have that
for a flat FRW model the energy density takes the form
$\rho(t)=\rho_0 a^{-2}$, i.e. $E \sim a^{-1}$  and the scale factor
is given by $a(t)=a_0 t$, where $a_0$ is a constant.

Notice that in general, the Maxwell electromagnetic tensor has only
three independent components, two for the vector electric field
$\vec{E}=(E_1,E_2)$ and one for the magnetic field $B$. Thus, in 2+1
dimensions only the electric component is inherently anisotropic. If
$E_1=E_2=0$ the magnetic component behaves like a perfect fluid with
an equation of state of the stiff matter. Thus, in 2+1 dimensions,
the equation of state~(\ref{2+1 B}) with $\rho(t)=\rho_0 a^{-4}$, $B
\sim a^{-2}$ and $a(t)=a_0 t^{1/2}$ provide the general solution for
flat FRW cosmologies sourced by a magnetic field.

From Eq.~(\ref{11}) the standard state equation $p=\rho/2$ is
obtained only if one considers a sum of plane waves, i.e. incoherent
isotropic black body radiation, where we have vacuum transverse
electromagnetic waves with $E=B$. Thus, as well as we have in 3+1
dimensions for incoherent isotropic black body radiation, in 2+1
dimensions the radiation equation of state~(\ref{stateeq3}) is also
fulfilled for plane waves. However, in 3+1 dimensions we have that
the radiation equation of state~(\ref{radiacion3+1}) is also
fulfilled for $B \neq E=0$, $E \neq B=0$; and $E \neq B$ with
non-vanishing magnetic and electric fields, thus the whole
parallelism with the four-dimensional case does not extend to
Maxwell electromagnetic fields satisfying the relation $E \neq B$ in
three-dimensional gravity.

\subsection{(2+1)-FRW cosmologies with a mixture of non-interacting electric and magnetic fields}

In the framework of FRW spacetimes the three-dimensional Maxwell
electromagnetic field may be interpreted as a cosmological
configuration with a mixture of two barotropic perfect fluids: a
matter component
\begin{eqnarray}\label{QEQ1B}
\rho_1(t)=\frac{B(t)^2}{2}
\end{eqnarray}
with a stiff equation of state ($p_1=\rho_1$) representing the
magnetic field, and a fluid
\begin{eqnarray}\label{EEQ2E}
\rho_2(t)=\frac{E(t)^2}{2}
\end{eqnarray}
with a dust equation of state ($p_2=0$) representing the electric
field. In this case both components  $\rho_1$ and $\rho_2$ satisfy
the conservation equation
\begin{eqnarray}\label{ConsEqEB}
\dot{\rho}_1+\dot{\rho}_2+2 \frac{\dot{a}}{a} (\rho_1+\rho_2 +p_1
+p_2 )=0,
\end{eqnarray}
implying that the sum of two fluids is conserved.

In order to find solutions, formally we can consider scenarios where
the electric field does not interact with the magnetic field, and
scenarios where the electric and magnetic fields interact with each
other.

Let us now consider the Maxwell equations for the studied
gravitational configuration. For the metric~(\ref{2+1FRW}) we may
write
\begin{eqnarray}
F=E_1 \theta^{(1)} \wedge \theta^{(0)} + E_2 \theta^{(2)} \wedge
\theta^{(0)}+B \theta^{(1)} \wedge \theta^{(2)},
\end{eqnarray}
where we have introduced the proper orthonormal basis
$\theta^{(0)}=dt$, $\theta^{(1)}=a(t)/\sqrt{1-k r^2} \, dr$ and
$\theta^{(2)} =a(t) \, r d \theta$. Thus, the Maxwell tensor, in the
coordinate basis, takes the form
\[F^{\mu \nu}= \left( \begin{array}{ccc}
0 & \frac{(1-k r^2)E_1}{a \sqrt{1-kr^2}} & \frac{E_2}{r a} \\
-\frac{(1-k r^2)E_1}{a \sqrt{1-kr^2}} & 0 & \frac{(1-k r^2)B}{r a^2 \sqrt{1-kr^2}} \\
-\frac{E_2}{r a} & -\frac{(1-k r^2)B}{r a^2 \sqrt{1-kr^2}} & 0
\end{array} \right),\]
and the Maxwell equations $F^{\mu \nu}_{\,\,\,\,\,\, ; \nu}=j^\mu$
and $F_{\alpha \nu; \mu}+F_{\mu \alpha; \nu}+F_{\nu \mu; \alpha}=0$
are respectively given by (the Greek indices run from $0$ to $2$)

\begin{eqnarray}\label{ME1}
F^{\mu \nu}_{\,\,\,\,\,\,  ; \nu}= \left( \begin{array}{c}
  \frac{\sqrt{1-k r^2}}{r a}  E_1  \\
  -\frac{\sqrt{1-k r^2}}{a^2} (a \dot{E_1}+E_1 \dot{a})   \\
-\frac{1}{r a^2} (a \dot{E_2}+E_2 \dot{a})
\end{array} \right)=j^\mu,
\end{eqnarray}
\begin{eqnarray}\label{ME2}
E_2+\frac{r}{\sqrt{1-k r^2}}(a \dot{B}+2 B \dot{a})=0.
\end{eqnarray}

Let us first consider the case of a vanishing electric field. We
obtain from Eq.~(\ref{ME1}) that $j^\mu=0$, while Eq.~(\ref{ME2})
implies that $a \dot{B}+2 B \dot{a}=0$, which is consistent with the
homogeneity and isotropy of the FRW metric. Hence, the magnetic
field is given by $B(t)=B_0/a^2(t)$, in agreement with what we have
stated above, in the previous section, for the 2+1 FRW magnetic
solution.

We consider next the inclusion of the electric field into the study.
It becomes clear that its vector character breaks the isotropy and
homogeneity symmetries of the FRW spacetimes. For non-vanishing
electric and magnetic fields, the inhomogeneous character of
Eq.~(\ref{ME2}) requires first that
\begin{eqnarray}
 \dot{B}+2 B \frac{\dot{a}}{a}=\frac{1}{2}
\frac{d}{dt} \left(B^2 \right)+ 2\frac{\dot{a}}{a} B^2=0,
\end{eqnarray}
and second that the electric component $E_2=0$. Consequently in
Eq.~(\ref{ME1}) $F^{2 \nu}_{\,\,\,\,\,\,  ; \nu}=0=j^2$. Now, by
taking into account Eq.~(\ref{QEQ1B}) and that
\begin{eqnarray}\label{unoF}
\dot{\rho}_1+2 \frac{\dot{a}}{a} (\rho_1 +p_1 )=\frac{1}{2}
\frac{d}{dt} \left(B^2 \right)+ 2\frac{\dot{a}}{a} B^2=0,
\end{eqnarray}
from Eqs.~(\ref{EEQ2E}) and~(\ref{ConsEqEB}) we obtain that
\begin{eqnarray} \label{dosF}
\dot{\rho}_2+2 \frac{\dot{a}}{a} (\rho_2 +p_2 )=\frac{1}{2}
\frac{d}{dt} \left(E^2 \right)+ \frac{\dot{a}}{a} E^2=0.
\end{eqnarray}
These two equations indicate us that  the electric and magnetic
fields are conserved separately, and hence the conservation
equation~(\ref{ConsEqEB}) is fulfilled.

Notice that due to that $E_2=0$ the RHS of Eq.~(\ref{dosF}) implies
that
\begin{eqnarray}\label{ECdeE15}
a \dot{E_1}+E_1 \dot{a}=0,
\end{eqnarray}
then in Eq.~(\ref{ME1}) we have that $F^{1 \nu}_{\,\,\,\,\,\,  ;
\nu}=0=j^1$. Thus, the radial electric field takes the form
$E_1=E_0/a$. Since, for time dependent electric and magnetic fields,
the Maxwell equations impose a restriction only on the non-radial
electric component $E_2$, which must vanish, then the Lorentz
invariant is given by $F/2=E^2-B^2=E_1^2-B^2$. We can have pure
electric field for $B^2=0$ or pure magnetic field for $E^2_1=0$, as
well as a mixture of both fields.

The presence of the radial electric component $E_1$ clearly breaks
the symmetries of the FRW spacetime. In order to fulfill them we
conclude that an average procedure must be applied for the radial
electric component $E_1$, and consequently in Eq.~(\ref{ME1}) $F^{0
\nu}_{\,\,\,\,\,\, ; \nu}=0=j^0$.

However, it must be remarked that this scenario requires that
$\overline{E^2_2}=0$, therefore it is not related to the
requirements of the spatial averaging procedure defined before in
Eqs.~(\ref{8}) and~(\ref{9}), since this one requires that
$\overline{E^2_1}=\overline{E^2_2}=E^2/2$, and
$E^2=\overline{E^2_1}+\overline{E^2_2}$, for a non-vanishing
electric field. In this case the Maxwell equations are trivially
fulfilled since for a such spatial averaging procedure we have that
$\overline{E_1}=\overline{E_2}=0$, as we can see from Eq.~(\ref{8}).
In what follows we shall discuss FRW solutions fulfilling the
average procedure defined in Eqs.~(\ref{8}) and~(\ref{9}).

From Eqs.~(\ref{unoF}) and~(\ref{dosF}) we conclude that for
non-interacting electric and magnetic fields we have that
\begin{eqnarray}\label{deB}
B(t)&=& \frac{B_0}{a^2(t)}, \\ \label{deE} E(t)&=& \frac{E_0}{a(t)},
\end{eqnarray}
respectively. The Friedmann equation~(\ref{2+1FE}) in this case
takes the following form:
\begin{eqnarray}\label{FREQB}
2 \left(\frac{\dot{a}}{a} \right)^2 = \kappa_3 \left(
\frac{B_0^2}{a^4}+\frac{E_0^2-2k/\kappa_3}{a^2} \right).
\end{eqnarray}

It becomes clear, for example, that for a flat FRW cosmology at
early times the magnetic component dominates over the electric
field, while for late times the electric field dominates over the
magnetic field. Note that from Eq.~(\ref{FREQB}) we have that the
general form of the scale factor is given by
\begin{eqnarray}\label{a sin interaction}
a^2(t)=\frac{\kappa_3 (E_0^2-2k/\kappa_3)}{2}
(C+t)^2-\frac{B^2_0}{(E_0^2-2k/\kappa_3)},
\end{eqnarray}
where $C$ is a constant of integration.

For $B \neq 0$ and $E \neq 0$ the invariant $F/2$ is given by
\begin{eqnarray}
B^2-E^2 = \frac{B_0^2}{a^4}-\frac{E_0^2}{a^2},
\end{eqnarray}
then if $0<a \leq B_0/E_0$, $B^2 \geq E^2$, while if $B_0/E_0 \leq a
< \infty$, $B^2 \leq E^2$.

It is useful to remark that, in the case of three-dimensional static
Einstein-Maxwell spacetimes, there exist the 2+1-analog of the
magnetic Reissner-Norsdtr\"{o}m spacetime, and separately the
electric Reissner-Norsdtr\"{o}m analog~\cite{Cataldo15C}. It is
noteworthy that the 2+1-magnetic Reissner-Nordstr\"{o}m analog is
not a black hole in contrast with the 2+1-electric
Reissner-Nordstr\"{o}m analog, where a black hole is
present~\cite{Cataldo15C}.

In order to close this subsection, we want to make some comments on
the possibility of considering (2+1)-FRW cosmologies with a mixture
of interacting electric and magnetic fields. In principle one can
introduce more general scenarios where the magnetic and electric
fields do not conserve separately and are coupled to each other. One
coupling mechanism can be formally introduced into the Friedmann
equations by defining an homogeneous interacting term Q(t) in the
following form~\cite{Cataldo15}:
\begin{eqnarray}\label{CEQ1}
\dot{\rho}_1+2 \frac{\dot{a}}{a} (\rho_1 +p_1 ) &=& Q(t), \label{CE1}\\
\label{CEQ2} \dot{\rho}_2+2 \frac{\dot{a}}{a} (\rho_2 +p_2 ) &=&
-Q(t), \label{CE2}
\end{eqnarray}
In this case $Q > 0$ is interpreted as a transfer of energy from
fluid $\rho_2$ to fluid $\rho_1$, while for  $Q < 0$, we should have
an energy transfer from fluid $\rho_1$ to fluid $\rho_2$. With the
help of Eqs.~(\ref{QEQ1B}) and~(\ref{EEQ2E}), and by taking into
account that $p_1=\rho_1$ and $p_2=0$, Eqs.~(\ref{CEQ1})
and~(\ref{CEQ2}) may be rewritten in the form
\begin{eqnarray}\label{QunoF}
\frac{1}{2} \frac{d}{dt} \left(B^2 \right)+ 2\frac{\dot{a}}{a}
B^2=Q(t),
\\
\label{QdosF} \frac{1}{2} \frac{d}{dt} \left(E^2 \right)+
\frac{\dot{a}}{a} E^2=-Q(t).
\end{eqnarray}
The interpretation of these equations is direct: for the case $Q >
0$ we have a transfer of energy from the electric field $E$ to the
magnetic field $B$, while if $Q < 0$, we should have an energy
transfer from the magnetic to the electric fields. Notice that
Eqs.~(\ref{QunoF}) and~(\ref{QdosF}) imply that the whole
conservation equation~(\ref{ConsEqEB}) is satisfied.

Nevertheless, in this case the inhomogeneous character of
Eq.~(\ref{ME2}) requires that $a \dot{B}+2 B \dot{a}=0$, implying
that $Q(t)=0$, so we are not allowed to consider such interacting
scenarios for time-dependent electric and magnetic fields in the
framework of homogeneous and isotropic cosmologies. However, it must
be noticed that, in principle, such an interaction between electric
and magnetic fields can be appropriately introduced in the framework
of more general metrics than FRW ones, such as for example
inhomogeneous circularly symmetric spacetimes depending, as well as
the electric and magnetic fields, on the time and radial
coordinates. For these interacting models the interacting term must
be considered in the form $Q=Q(t,r)$. This work is currently in
progress.

\subsection{Three-dimensional conformally invariant electrodynamics}
In 3+1-dimensions the equation of state $p=\rho/3$ is a direct
consequence of the conformal invariant character of the Maxwell
equations. Effectively, it can be shown that Maxwell equations in
four dimensions are invariant under conformal transformation
$\tilde{g}_{\alpha \beta}=\Omega^2 g_{\alpha \beta}$ and
$\tilde{F}_{\mu \nu}=F_{\mu \nu}$~\cite{Wald}. The conformal
invariance of these equations is encoded by the traceless condition
$T=T_{\mu \nu} g^{\mu \nu}=0$ of the energy-momentum
tensor~(\ref{TEMEM}) in 3+1-dimensions. Thus, the Maxwell field in
four dimensions has conformal symmetry. This result is true
regardless of whether spacetime is flat or curved. In spacetime
dimensions with $N \neq 3$ this is not true anymore because the
Maxwell energy-momentum tensor possesses a non-vanishing
trace~\cite{Wald}.

Fortunately, we can take advantage of this conformal symmetry by
using an extension of the Maxwell action that possesses the
conformal invariance in an arbitrary dimension. The Maxwell action
in N+1-dimensions may be written as~\cite{Mokhtar}
\begin{eqnarray}\label{action MM}
S_M =\alpha \int  \sqrt{-g} \, \left(F_{\mu \nu} F^{\mu \nu}
\right)^\frac{N+1}{4} \, d^{N+1}x,
\end{eqnarray}
where $F_{\mu \nu}= \partial_\mu A_\nu-\partial_\nu A_\mu$. It is
simple to see that under a conformal transformation acting on the
metric and the electromagnetic fields as $\tilde{g}_{\alpha
\beta}=\Omega^2 g_{\alpha \beta}$ and $A_\mu \rightarrow A_\mu$,
this action remains unchanged~\cite{Mokhtar}. The energy-momentum
tensor associated to the action~(\ref{action MM}) is given by
\begin{eqnarray}\label{TEMEM MM}
T_{\alpha \beta}=4\alpha \left(F_{\gamma
\delta} \, F^{\gamma \delta} \right)^{\frac{N+1}{4}} \times \nonumber \\
\left[ -\frac{N+1}{4} F_{\alpha \gamma}F_{\beta}^{\,\,\, \gamma}
\left(F_{\gamma \delta} \, F^{\gamma \delta} \right)^{-1} +
\frac{1}{4} g_{\alpha \beta} \right].
\end{eqnarray}
It can be shown that the traceless condition for energy-momentum
tensor~(\ref{TEMEM MM}) is fulfilled.

The Maxwell action~(\ref{action MM}) in 2+1-dimensions takes the
form
\begin{eqnarray}\label{action MM 2+1}
S_M =\alpha \int  \sqrt{-g} \, (F_{\mu \nu} F^{\mu \nu})^{3/4} \,
d^3x,
\end{eqnarray}
hence the energy-momentum tensor associated to the
action~(\ref{action MM 2+1}) is given by
\begin{eqnarray}\label{TEMEM MM 2+1}
T_{\alpha \beta}=3 \alpha \left(F_{\gamma \delta} \, F^{\gamma
\delta} \right)^{-\frac{1}{4}} \left[-  F_{\alpha
\gamma}F_{\beta}^{\,\,\,\,\, \gamma}  + \frac{1}{3} \, g_{\alpha
\beta} \, \left(F_{\gamma \delta} \, F^{\gamma \delta} \right)
\right]. \nonumber \\
\end{eqnarray}
It is interesting to note that this 2+1-electrodynamics is a
nonlinear electrodynamics. Such three-dimensional nonlinear
electrodynamics was discussed before in the
literature~\cite{Cataldo,Cataldo15A,Cataldo15B,Nonlinear}. In
general, one can construct a (2+1)-Einstein theory coupled with
nonlinear electrodynamics starting from the action
\begin{eqnarray}\label{nonlinear action 2+1}
S_{NL}=\int \sqrt{-g} \,  L(F) \, dx^3,
\end{eqnarray}
where the electromagnetic Lagrangian $L(F)$ depends upon a single
invariant
\begin{eqnarray}\label{invariante}
F=\frac{1}{4} F^{\mu \nu} \, F_{\mu \nu}=\frac{1}{2} \, (B^2-E^2).
\end{eqnarray}
Physically one requires the Lagrangian to coincide with the linear
Maxwell $L(F)=-F/4 \pi$ at small values of the electromagnetic
fields. The energy-momentum tensor associated to
action~(\ref{nonlinear action 2+1}) is given by
\begin{eqnarray}\label{Tmnnolin}
T_{\mu \nu} = g_{\mu \nu} L(F) - F_{\mu \gamma} \, F_{\nu}^{\,\,\,\,
\gamma} L_{_{,F}},
\end{eqnarray}
where $L_{,F}$ denotes the derivative of L(F) with respect to $F$.
The trace of this tensor is given by
\begin{eqnarray}
T=3 \, L(F)-4 \, F L_{_{,F}},
\end{eqnarray}
therefore, by requiring $T$ to vanish, we establish the existence of
the unique 2+1-nonlinear electrodynamics, with vanishing
energy-momentum trace, given by the action~(\ref{action MM 2+1}).
This nonlinear electrodynamics was considered first for obtaining a
(2+1)-dimensional static black hole with Coulomb-like
field~\cite{Cataldo}.

Thus, the conformally invariant 2+1-electrodynamics~(\ref{action MM
2+1}) is a particular case of three-dimensional nonlinear
electrodynamics described by the action~(\ref{nonlinear action
2+1}). The same can be said about any higher dimension as well.
Indeed, any N+1-electrodynamics described by the action~(\ref{action
MM}) is a particular case of N+1-nonlinear electrodynamic theories,
characterized by having a traceless energy-momentum tensor, and
hence by being conformally invariant.

The averaging procedure, applied to the electric component
$E_{_{i}}$ and the magnetic field $B$, yields relations ~(\ref{8})
and~(\ref{9}), namely, for the metric $(+1,-1,-1)$:
$\overline{E_i}=\overline{E_i \, B}=0$, $\overline{E^2_i}=E^2/2$,
$\overline{B^2}=B^2$.

Consequently the average of the energy-momentum tensor of the
nonlinear electrodynamics under consideration gives rise to the
relations
\begin{eqnarray}
T^{00}_{_{2+1}}&=&\rho^{_{nonlE}}_{_{2+1}},
T^{11}_{_{2+1}}=T^{22}_{_{2+1}}= p^{_{nonlE}}_{_{2+1}},
\end{eqnarray}

\begin{eqnarray}
\label{pnonlE} p^{_{nonlE}}=\frac{1}{2} \, \rho^{_{nonlE}} =
\frac{\alpha(E^2+2 B^2)}{2 \left |4 F^{\gamma \delta} F_{\gamma
\delta} \right|^{1/4} }.
\end{eqnarray}
as one should expect. Note that Eqs.~(\ref{invariante})
and~(\ref{pnonlE}) imply that in this specific electrodynamics we
may consider only cases with $E \neq B$ in order to have finite
energy density and pressure in Eq.~(\ref{pnonlE}).

Therefore, we conclude that if one considers a three-dimensional
perfect fluid with the radiation equation of state $p=\rho/2$, this
must be done in the framework of the nonlinear conformally invariant
electromagnetic theory described by the action~(\ref{action MM 2+1})
and energy-momentum tensor~(\ref{TEMEM MM 2+1}).

Cornish and Frankel~\cite{Cornish} derived, among others, a
cosmological solution, using the (2+1)-FRW metric, which fulfills
the law~(\ref{stateeq3}), referring to it as radiation-dominated FRW
universe, see Eqs. (4.1)-(4.7) of the quoted work. On the light of
the present results, the Cornish and Frankel solution (see also the
Ref.~\cite{Barrow}) has to be associated, from the point of view of
electrodynamics, to plane waves in the framework of the linear
Maxwell electrodynamics, and for more general cases with $E \neq B$
to the nonlinear conformally invariant electrodynamics exhibited
above.

\section{acknowledgements}
We thank the anonymous Referee for the careful reading of our
manuscript and the valuable comments. M.C. acknowledges the
hospitality of the Physics Department of CINVESTAV were this work
was in part done. This work was also supported in part by Grant
CONICYT-CONACYT 2001-5-02-159 (M.C. and A.G.), CONICYT through grant
FONDECYT N$^0$ 1080530 (M.C.). It also was supported by the
Direcci\'on de Investigaci\'on de la Universidad del Bio-B\'\i o
through the grant N$^0$ DIUBB 121007 2/R and N$^0$ GI121407/VBC
(M.C.).


\begin{thebibliography}{xxxxx}
\bibitem{Landau} L. D. Landau et al, Statistical Physics
(Butterworth-Heinemann, Oxford, 1999); L.D. Landau and E.M.
Lifshitz, The Classical Theory of Fields (Butterworth-Heinemann,
Oxford, 1997).
\bibitem{Lightman} A. Lightman et al, Problem Book in Relativity and
Gravitation (Princeton University Press, Princeton, New Jersey,
1979).
\bibitem{Cataldo} M. Cataldo, N. Cruz, S. del Campo and A.
Garc\'\i a, Phys. Lett. B {\bf 484}, 154 (2000).
\bibitem{Tolman} R. Tolman and P. Ehrenfest, Phys. Rev. {\bf
36}, 1791 (1930).
\bibitem{Enqvist} K. Enqvist and P. Olesen, Phys. Lett. B, {\bf
319}, 178 (1993); M.~Hindmarsh and A.~Everett,
  Phys.\ Rev.\ D {\bf 58}, 103505 (1998). 
\bibitem{Novello} V.A. De Lorenci, R. Klipert, M. Novello and J.M. Salim,
Phys. Rev. D {\bf 65}, 063501 (2002).
\bibitem{Cataldo15C} M.~Cataldo, J.~Crisostomo, S.~del Campo and P.~Salgado,
  Physics Letters B {\bf 584}, 123 (2004); 
M.~Cataldo and P.~Salgado,
  Phys.\ Rev.\ D {\bf 54}, 2971 (1996); 
E.~W.~Hirschmann and D.~L.~Welch,
  Phys.\ Rev.\ D {\bf 53}, 5579 (1996); 
G. Clement, Class. Quant. Grav. {\bf 10}, L49 (1993); P. Peld\'an,
Nucl. Phys. B {\bf 395}, 239 (1993); M. Ba\~nados, C. Teitelboim and
J. Zanelli, Phys. Rev. Lett. {\bf 69}, 1849 (1992).
\bibitem{Cataldo15} R.~Ghosh, S.~Chattopadhyay and U.~Debnath,
  Int.\ J.\ Theor.\ Phys.\  {\bf 51}, 589 (2012); M.~U.~Farooq, M.~Jamil and U.~Debnath,
  Astrophys.\ Space Sci.\  {\bf 334}, 243 (2011); H.~Farajollahi, N.~Mohamadi and H.~Amiri,
  Mod.\ Phys.\ Lett.\ A {\bf 25}, 2579 (2010); M.~Cataldo, F.~Arevalo and P.~Minning,
  JCAP {\bf 1002}, 024 (2010).
\bibitem{Wald} R. Wald, General Relativity, The University of
Chicago Press, 1984.
\bibitem{Mokhtar} M.~Hassaine and C.~Martinez,
  Phys.\ Rev.\ D {\bf 75}, 027502 (2007). 
\bibitem{Cataldo15A} M.~Cataldo and A.~Garcia,
  Phys.\ Rev.\ D {\bf 61}, 084003 (2000).
\bibitem{Cataldo15B} M.~Cataldo and A.~Garcia,
  Phys.\ Lett.\ B {\bf 456}, 28 (1999).
\bibitem{Nonlinear} S.~Gangopadhyay and D.~Roychowdhury,
  JHEP {\bf 1205}, 002 (2012); S.~H.~Mazharimousavi, O.~Gurtug, M.~Halilsoy and O.~Unver,
  Phys.\ Rev.\ D {\bf 84}, 124021 (2011); L.~Balart,
  Mod.\ Phys.\ Lett.\ A {\bf 24}, 2777 (2009);  T.~K.~Dey,
  Phys.\ Lett.\ B {\bf 595}, 484 (2004); R.~Yamazaki and D.~Ida,
  Phys.\ Rev.\ D {\bf 64}, 024009 (2001).
\bibitem{Cornish} N. Cornish and N. Frankel, Phys. Rev. D {\bf
43}, 2555 (1991).
\bibitem{Barrow} J.~D.~Barrow, D.~J.~Shaw and C.~G.~Tsagas,
  Class.\ Quant.\ Grav.\  {\bf 23}, 5291 (2006).
\end{thebibliography}
\end{document}